# Transient lensing from a photoemitted electron gas imaged by ultrafast electron microscopy

*Omid Zandi[1,2], Allan E. Sykes[1,2], Ryan D. Cornelius[1,2], Francis M. Alcorn[1,2], Brandon Zerbe[4], Phillip M. Duxbury[4], Bryan W. Reed[5], Renske M. van der Veen[1,2,3]*

[1] Department of Chemistry, University of Illinois at Urbana-Champaign, Urbana, IL 61801, USA

[2] Materials Research Laboratory, University of Illinois at Urbana-Champaign, Urbana, IL 61801, USA

[3] Department of Materials Science and Engineering, University of Illinois at Urbana-Champaign, Urbana, IL 61801, USA

[4] Department of Physics and Astronomy, Michigan State University, East Lansing, MI 48824, USA

[5] Integrated Dynamic Electron Solutions, Inc. (IDES), Pleasanton, CA 94588, USA

ABSTRACT

Understanding and controlling ultrafast non-equilibrium charge carrier dynamics is of fundamental importance in diverse fields of (quantum) science and technology. Here, we create a three-dimensional hot electron gas through two-photon photoemission from a copper surface in vacuum. We employ an ultrafast electron microscope to record movies of the subsequent electron dynamics on the picosecond-nanosecond time scale. After a prompt Coulomb explosion, the subsequent dynamics is characterized by a rapid oblate-to-prolate shape transformation of the electron gas, and exceptionally periodic and long-lived electron cyclotron oscillations inside the magnetic field of the objective lens. In this regime, the collective behavior of the oscillating electrons causes a transient, mean-field lensing effect and pronounced distortions in the images. We derive an analytical expression for the time-dependent focal length of the electron-gas lens, and perform numerical electron dynamics and probe image simulations to determine the role of Coulomb self-fields and image charges. This work paves the way for the direct visualization of cyclotron oscillations inside two-dimensional electron-gas materials and the elucidation of electron/plasma dynamics and properties that could benefit the development of high-brightness electron and X-ray sources.

INTRODUCTION

Understanding the non-equilibrium dynamics of charge carriers (electrons/ions/holes) is of uttermost importance in a vast range of fundamental and technological fields, including chemistry, solid-state physics, plasma physics, and high-brightness electron sources. Carrier motion often unfolds on ultrafast time scales and requires tools that can directly visualize the dynamics with appropriate spatial and temporal resolutions, *i.e.* Ångstroms-micrometers (Å-μm) and femtoseconds-nanoseconds (fs-ns), respectively. In this regard, ultrafast electron microscopy (UEM) has recently emerged as a powerful technique for the study of ultrafast photoinduced processes in nanoscale systems[1-14]. The material is excited by a short fs-ns laser pulse, which is followed by a similarly short electron pulse that probes the ensuing dynamics by means of imaging, diffraction, or spectroscopy inside a transmission electron microscope (TEM).

Here, we use UEM to visualize the ultrafast evolution of a hot three-dimensional (3D) photoemitted electron gas under a static magnetic field in real time and real space. Confined electron gases[15] can exhibit intriguing properties such exceptionally high electron mobilities[16], quantum Hall effects[17,18], Shubnikov-de Haas oscillations[19], anomalous de Haas-van Alphen effects[20], and superradiant damping[21]. Understanding and controlling these phenomena is of fundamental importance in diverse fields of quantum science and technology[22,23]. For example, two-dimensional (2D) electron gases at semiconducting heterointerfaces or in 2D materials, that are subject to an external magnetic field, have been studied by frequency- and time-domain THz spectroscopies[17,21,24-26]. An electron gas in a uniform magnetic field executes circular Larmor orbits in a plane perpendicular to the magnetic field. Transitions between the eigenstates (Landau levels) of electron gases confined by a magnetic field are called cyclotron resonances, whose frequencies, line widths, and decays have been used to determine band structures, effective masses, carrier densities, mobilities and scattering times in semiconducting materials[21,25,27-30]. Quantum effects arising from Landau levels are dominant when the mean thermal energy of the gas is smaller than the energy level separation, which means experiments are often performed at low temperatures and under strong magnetic fields.

The proof-of-principle UEM experiments on 3D electron gases in uniform magnetic fields presented in this work pave the way for the direct visualization of cyclotron oscillations inside materials, in particular 2D electron-gas systems such as GaAs/AlGaAs[18,25,31] or graphene[26,32,33]. In contrast to frequency- or time-domain THz/microwave spectroscopic investigations, performing such experiments inside an ultrafast electron microscope enables spatially resolving photoexcited electron density variations, similar to previous scanning probe microscopy experiments[31,34-36] but with fs-picosecond (ps) temporal resolution. Furthermore, the capability to image and temporally resolve photoemitted carriers is highly relevant in the plasma physics community[37-40], and for the development and characterization of high-brightness electron sources for fourth-generation X-ray facilities or ultrafast electron diffraction and microscopy setups[41-49]. The analytic model we develop here allows rough approximation of the number of electrons in the photoemitted gas, which is directly correlated with the electron lens magnification, as well as their velocity spread. Systematic variations of the laser fluence and wavelength, and adding a bias to the sample, will enable obtaining valuable insight into the electron emission process and the subsequent processes that affect electron beam properties such as emittance.

RESULTS

**Direct imaging of electron cyclotron oscillations on the picosecond time scale.** We performed our experiments using a modified environmental TEM operating at 300 keV (Fig. 1a), which is interfaced with a high repetition-rate, fs laser system (see Methods section for more details). Laser pump pulses (~200 fs, 528 nm, ~33 mJ/cm$^2$) are guided onto the sample using a mirror/lens system that inserts into the energy-dispersive spectroscopy (EDS) port of the TEM. Short probe electron pulses (<1 ps) are generated by impinging a UV laser beam onto a LaB$_6$ photocathode. The laser pump and electron probe pulses are precisely synchronized in time at a repetition rate of 490 kHz and their relative delay is adjusted by means of an optical delay stage. In this way, we record real-space movies of the charge density dynamics after laser excitation with an integration time of 1 second ($5 \cdot 10^5$ shots) per frame. The sample consists of a tilted 3000 mesh copper grid with ~4.5 μm hexagonal holes, which are significantly smaller than the laser footprint of 22×36 μm (full width at half maximum, FWHM) on the sample. Since the photon energy (2.35 eV) is half the work function of copper (4.5-4.7 eV[50]), photoelectrons are emitted in the two-photon absorption regime[44,51]. The laser fluence is low enough that ablation and plasma formation can be discarded[52].

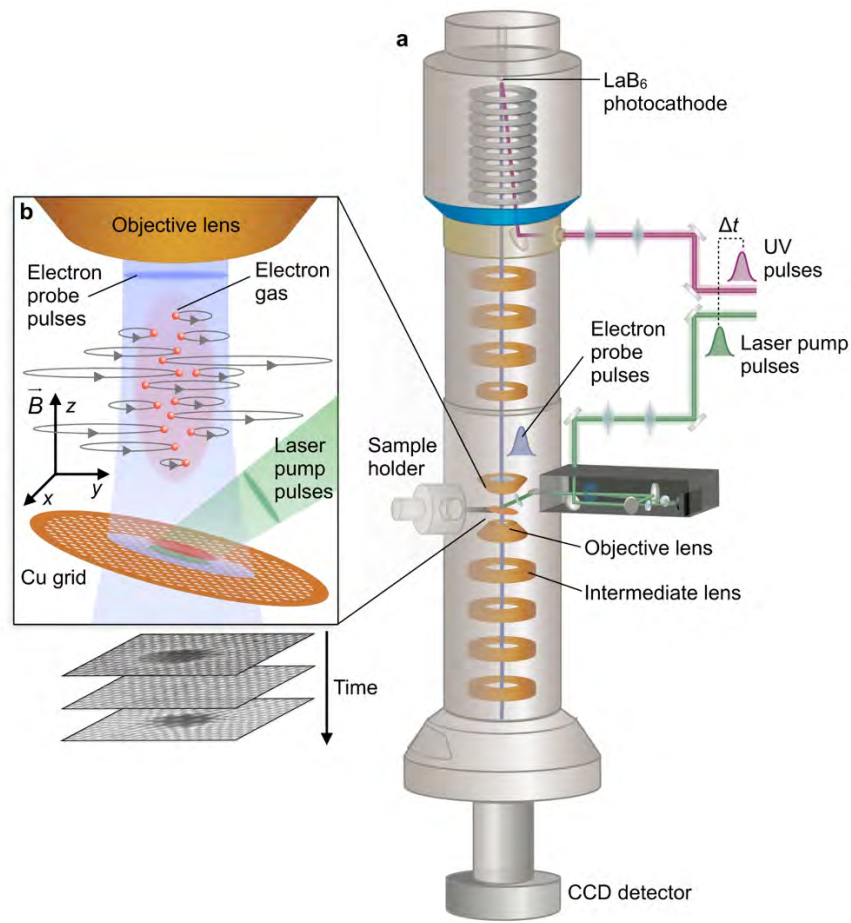

**Figure 1: Schematic of the ultrafast electron microscope used for imaging photoemitted electron gas dynamics. a** Short probe electron pulses (~500 fs, 500 kHz, 300 keV) are generated by impinging a UV (256 nm) laser beam onto a $LaB_6$ photocathode. Pump laser pulses (~200 fs, 528 nm, 490 kHz, ~33 mJ/cm$^2$) are focused onto the Cu grid sample inside a modified TEM. **b** A hot electron gas (red) is created by means of two-photon emission, which acts as a biconcave diverging lens to the probe electrons. After initial Coulomb explosion, the electron gas executes cyclotron oscillations (gray orbits) inside the magnetic field ($\vec{B}$) of the objective lens, which are resolved by changing the relative timing of the pump and probe pulses ($\Delta t$). The strength of the intermediate lens in the TEM can be tuned to obtain different imaging conditions. The sample is tilted by 15º towards the pump laser in order to minimize the laser footprint on the sample.

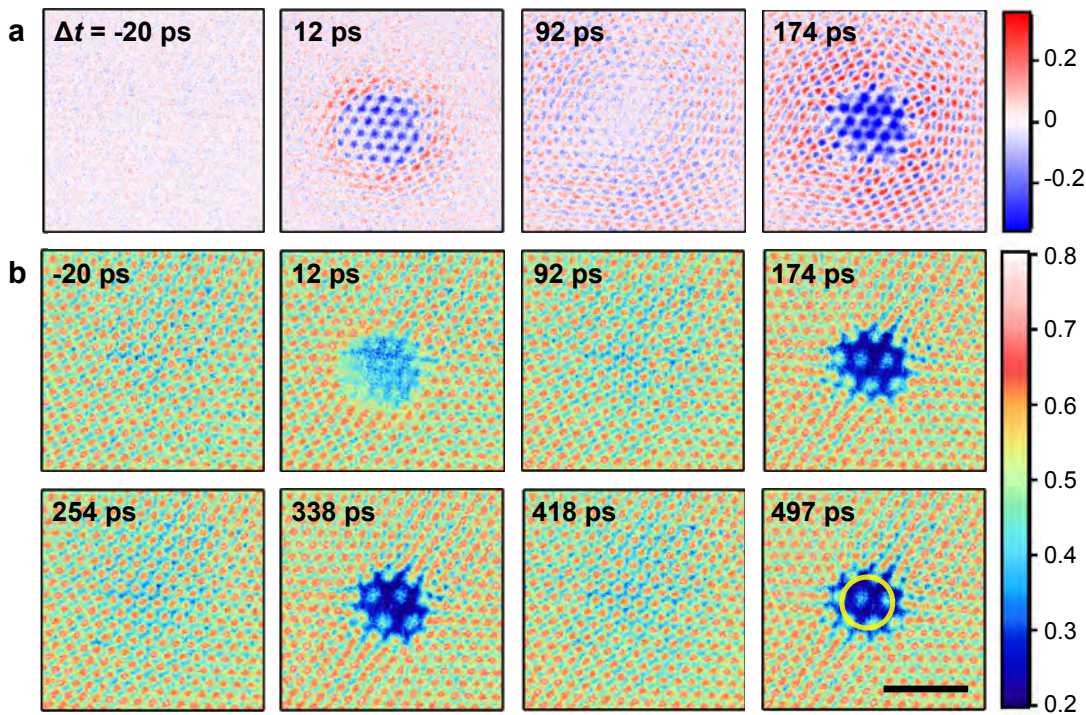

**Figure 2: Transient electron-gas lensing images.** Series of 3000 mesh copper grid images (**b**) and difference images (**a**) extracted from a ps-resolved UEM movie (528 nm, 200 fs, 33 mJ/cm$^2$ laser excitation) with an objective lens current of 0.7 A. The time delays correspond to the first few local maxima and minima in the ROI difference intensity trace in Fig. 3. The difference images were generated by subtracting an averaged image before time zero ($\Delta t = 0$). A typical region-of-interest (ROI) circle that is used to make plots of the intensity changes due to lensing is indicated in the last frame. The scale bar at the bottom right is 50 μm and applies to all images. The intermediate lens current was set to 0.65 A.

Fig. 2a,b show a set of copper grid (difference) images that is extracted from a ps-resolved UEM movie (see Supplementary Information, SI, movies S1 and S2). Upon laser excitation, a prompt depletion of intensity in the probe image occurs (12 ps frame), which lasts for approximately 50 ps. The subsequent dynamics is characterized by localized, periodic barrel distortions of the grid images at intervals of ~165 ps, with almost no changes in between resonance peaks. A region-of-interest (ROI) analysis for a similar data set (Fig. 3) shows that these recurring image distortions continue for more than twenty cycles over a time span of ~4 ns, and with a

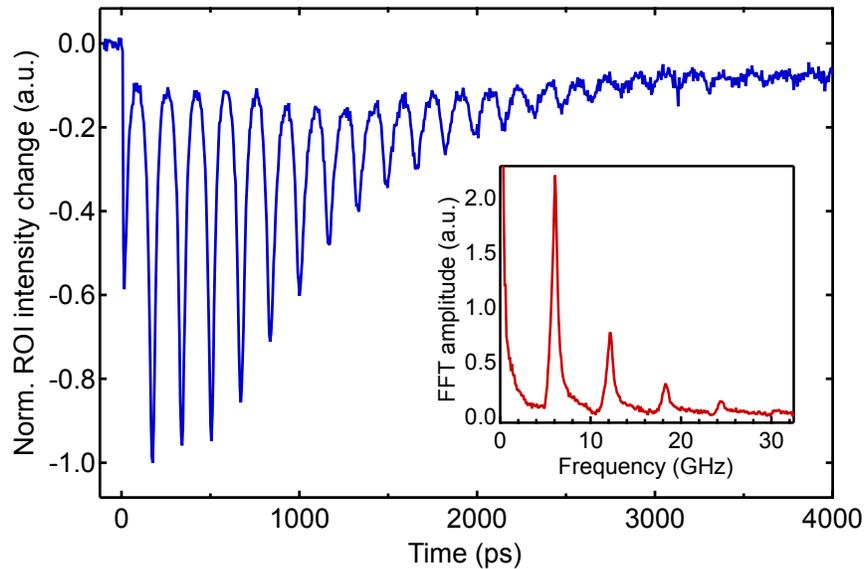

**Figure 3: Region-of-interest (ROI) intensity analysis of transient lensing dynamics.** ROI difference intensity (relative to before time zero) as a function of time delay. A typical ROI is denoted in Fig. 2b (different data set). The inset shows the FFT of the trace, with a fundamental frequency of 6.05 GHz (*i.e. T* = 165 ps). The objective lens current was set to 0.7 A and the intermediate lens current was set to 0.65 A.

fundamental frequency of 6.05 GHz as shown by the fast Fourier transformation (FFT) of the ROI trace. The oscillations are exceptionally periodic and long-lived. Except for higher harmonics resulting from the pointed shape of the resonance peaks, no other frequencies are contained in the data. The barrel distortions are due to a transient lensing effect of the photoemitted electron gas, whose transverse electric field causes a deflection of the probe electron pulses leading to a magnification in the projected image on the detector. The photoemitted electron cloud therefore acts as a 3D biconcave, diverging lens to the probe electron pulses.

The electron gas is subject to a static magnetic field ($\boldsymbol{B}$) along the electron optical axis which is imposed by the objective lens (upper and lower pole piece) of the TEM. This field is relatively weak due to the fact that we operate the TEM in low-magnification mode. Under the influence of the Lorentz force $\boldsymbol{F} = q\boldsymbol{v} \times \boldsymbol{B}$, the electrons gyrate around the magnetic field axis, with an electron cyclotron period of $T = 2\pi m_e/eB$, where $\boldsymbol{v}$ is the velocity vector of the electrons, $m_e$ is their mass, and $e$ is their charge (Fig. 1b). The (non-relativistic) radius of gyration for each electron is given by $r = v_T m_e/eB$, with $v_T$ the transverse $(x, y)$ velocity component in the direction perpendicular to the magnetic field $(z)$. Each electron therefore circulates with a different radius, depending on its initial velocity, but all electrons reconvene to their initial positions in the $x, y$-plane after a full cyclotron period $T$. Slightly before this point in time, the collective width of the electrons reaches a minimum and the transverse electric mean-field maximizes resulting in a pronounced transient lensing effect that is observed in the probe image. While the magnetic field confines the electron gas in the transverse direction, the longitudinal dynamics is affected by the $z$-velocity profile and the boundary conditions at the surface of the copper grid. We show later that this anisotropic confinement causes a rapid oblate-to-prolate shape transformation of the 3D electron gas and a large concurrent increase of the lensing strength on the time scale of ~100 ps. Fig. 4a shows ROI difference intensity traces recorded at various objective lens currents (OLC). Using the cyclotron period formula above and a FFT analysis of the ROI traces, we constructed a plot of the OLC versus cyclotron period and magnetic field (Fig. 4b). A linear relationship between OLC and magnetic field is obtained, which matches data from the TEM manufacturer (see SI1). The ROI intensity trace for OLC = 0 A (no magnetic field) merely shows the first peak, as expected.

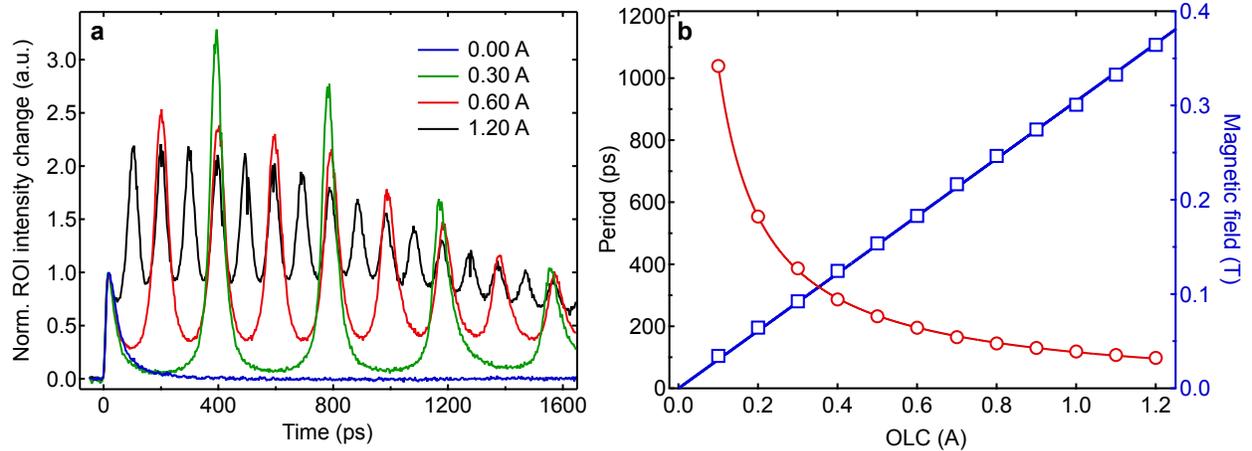

**Figure 4: Dependence on the objective lens current. a** ROI difference intensity traces as a function of objective lens current (OLC). The traces have been normalized to the height of the first peak ~12 ps after photoexcitation. The intermediate lens current was set to 1.1 A. Note that there are no oscillations for an OLC = 0 A. **b** Cyclotron oscillation period (red, left axis) as a function of the OLC. Using the cyclotron period formula $T = 2\pi m_e/eB$, the corresponding magnetic field at the sample position is calculated (blue, right axis). The latter corresponds well with data from the manufacturer (see SI S1). The solid blue line is a linear fit to the data.

We note that deflection effects due to transient electric fields from photocreated electron plumes have been observed previously in ultrafast electron diffraction and microscopy setups[37,38,53-59]. However, we report for the first time a detailed study of the space-charge dynamics in the presence of a magnetic field and the consequent changes in the image under various experimental conditions.

**Dependence on the imaging conditions and electron-gas astigmatism.** The transient lensing effects are only visible in the images if the projection lens system of the TEM is set to out-of-focus image conditions. This is a consequence of the geometry. Because the laser impinges on the grid from above, and because the fill fraction of the grid is quite high (66%), nearly all the photoemitted electrons will be above the grid. The grid itself will act as an electrostatic boundary condition preventing the space-charge electric fields from penetrating significantly below. Thus, the lensing effect of the electron cloud will deflect electrons radially before they strike the grid but will have essentially no effect on the grid pattern itself or on the focusing action produced by the TEM lenses below the grid. In a focused real-space image, the post-sample lenses map the $(x, y)$ spatial positions of electrons as they emerge from the back of the sample linearly onto the camera, suppressing information about the angles of the electron trajectories. Thus, an in-focus image should just produce an image of the grid, as we observe. To detect the space-charge lensing effect, we defocus the imaging system so that the resulting image is a linear combination of the spatial and angular coordinates of the electrons emerging from the back of the sample. Because the OLC is an extremely important parameter for the space-charge dynamics, we instead adjust the current of the first intermediate lens (IL), *i.e.* the first lens after the objective lens (see Fig. 1a). We recorded lensing movies (see movies S2-S4) for a range of IL excitation strengths. In this way, we are able to tune the electron-gas lensing effect from a magnifying, barrel image distortion for low IL excitations (Fig. 5a), to a demagnifying, pincushion image distortion for high IL excitations (Fig. 5c). For intermediate IL strength (Fig. 5b) we can image the post-sample crossover of the objective lens onto the detector. Thus, the defocused images reveal the shift in the post-sample crossover caused by the lensing effect of the electron gas, manifesting as a magnification or demagnification of the affected region relative to the rest of the grid.

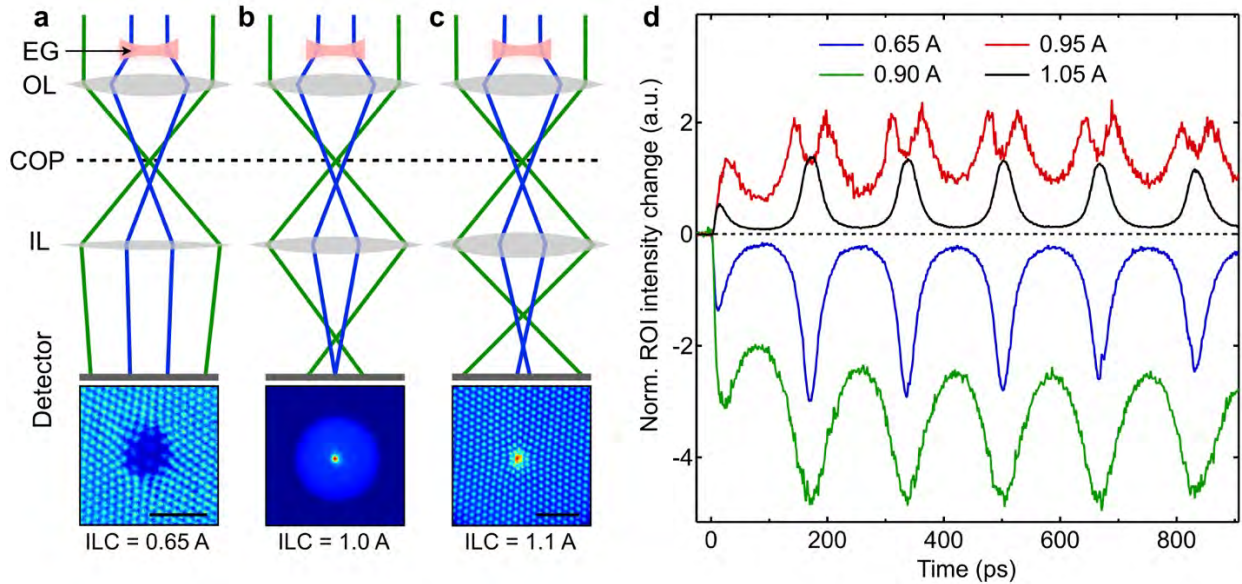

**Figure 5: Dependence on the intermediate lens strength. a-c** Image conditions with various intermediate lens currents (ILC): **a** 0.65 A, underfocus regime, barrel distortion, **b** 0.95 A, in-focus regime, **c** 1.1 A, overfocus regime, pincushion distortion. The corresponding images are shown at the bottom ($\Delta t$ = 173 ps, OL current 0.7 A). Since the electron gas (EG, red) acts like a diverging lens, the probe electrons that pass through the gas (blue) are focused by the objective lens (OL) after/below the probe electrons that are not affected by the electron lens (green). COP = cross-over plane. The scale bars are 50 μm (no scale bar for image in **b**, since the grid is not visible). In this figure, all the projection lenses after the sample are replaced by one equivalent lens (labeled IL). Only the IL strength was changed. The copper grid is not shown for simplicity. **d** ROI difference intensity traces for various ILC values (OL current 0.7 A).

Since the electron cloud acts like a diverging lens, the probe electrons that pass through the cloud are focused by the objective lens after/below the probe electrons that pass further away and are not affected by the electron gas. If the IL current (ILC) is set to a value such that a plane between the post-sample crossover of the unaffected probe electrons and the post-sample crossover of the lens-affected probe electrons is imaged onto the detector (Fig. 6), the ROI difference intensity trace shows a double-peak structure as shown in Fig. 5d for ILC = 0.95 A. A representative set of difference images recorded under such conditions is shown at the bottom of Fig. 6 (ILC = 0.95 A, OLC = 0.7 A), from which we see that the focal point is significantly astigmatic, *i.e.* the elongated shape of the spot rotates by approximately 90° as it moves through focus. We ascribe this astigmatism to the fact that the electrons are emitted from a slightly (15°) tilted grid surface. Their initial velocity distribution is aligned perpendicular to the sample surface, which translates into an increased transverse velocity component in the tilt direction relative to the electron-optical axis ($z$). In addition, the shape of the photoemitted electron gas is not perfectly circular in the $x,y$-plane due to shallow angle (37°) between the excitation laser and the sample surface, and the ensuing elliptical footprint of the laser on the sample. The astigmatism is not visible in all data sets, as it can be compensated by the condenser stigmator correction lenses inside the TEM.

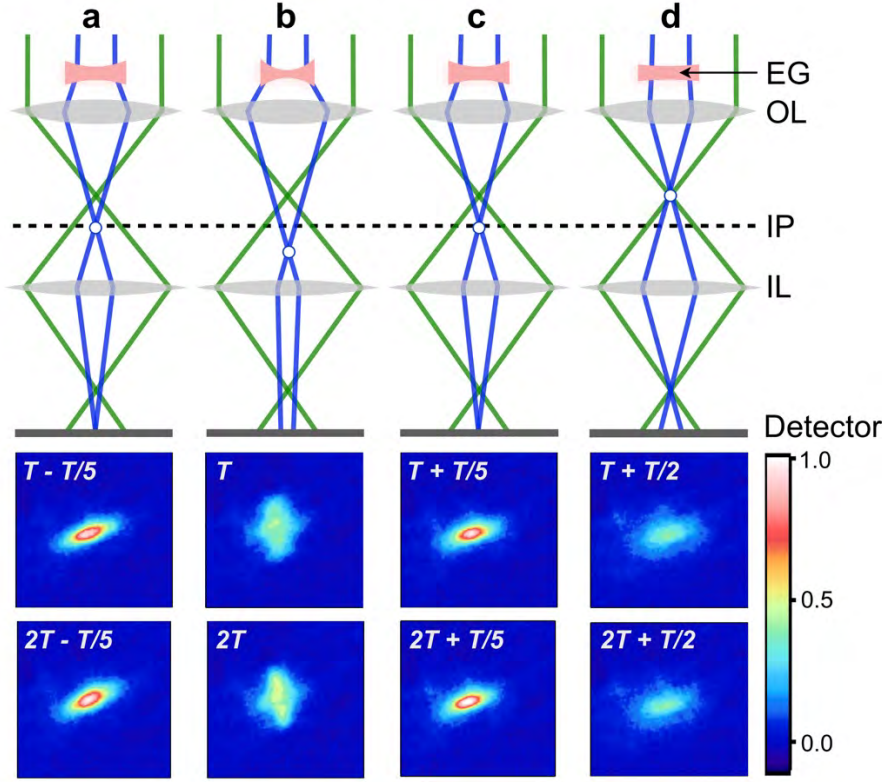

**Figure 6: Double-peak structure in ROI difference intensity traces and astigmatism of the electron-gas lens.** Schematics of the electron gas (EG, red) acting as a biconcave lens while it reaches its largest lensing strength (**b**). The intermediate lens (IL) current (0.95 A) is set such that the image plane (IP) denoted by the dashed line is projected onto the detector. Depending on the strength of the EG lens, the cross-over point (open circle) lies either on (**a,c**), below (**b**), or above (**d**) the IP. Under these conditions the ROI difference intensity shows a double-peak structure (see *e.g.* Fig. 5d, different data set), with local maxima appearing at approximately $nT \pm T/5$ ($n = 0,1,2,...$), $T$ being the cyclotron period as defined in the text. The focal point is astigmatic due to the elliptical shape of the EG lens, as is seen by the rotation of the elongated focal point as it moves through the IP. The objective lens (OL) current was set to 0.7 A, with $T$ = 165 ps.

**Analytical model for electron density evolution and determination of the transient focal length.** In order to quantitatively describe the evolution of the 3D electron gas, we developed an analytical model that allows us to estimate the velocity spread of the photoemitted electrons, the number of electrons in the gas, as well as the time-dependent focal length of the electron lens. It is known that the velocity profile of one-component plasma systems expanding under their mutual repulsion, known in the literature as Coulomb explosion, is largely determined by the early dynamics of the expanding bunch where the density, and hence the force between particles, is largest[42,60,61]. Therefore, we focus our analysis on the time range >50 ps once the velocity profile has largely been established (see below for numerical simulations that confirm this). In this regime we describe the electron gas as a cylindrically symmetric Gaussian charge distribution with a time-dependent transverse radius $\sigma_T(t)$, and an axial radius $\sigma_z(t)$ ($\sigma$ denotes standard deviation). While the evolution of $\sigma_z$ is essentially linear in time due to the almost free (linear) expansion of the gas in this direction, the transverse radius is affected by the Lorentz force $\boldsymbol{F} = q\boldsymbol{v} \times \boldsymbol{B}$. Solving the equations of motion (see SI2.1 for details), we obtain the periodic dependence for the transverse radius of the electron gas

$$\sigma_T(t) = \sqrt{2\left(\frac{\sigma_v}{\omega}\right)^2 (1 - \cos(\omega t)) + \sigma_r^2} \quad , \tag{1}$$

where $\omega = eB/2m_e$ is the cyclotron angular frequency, $\sigma_v$ is the velocity spread in the transverse direction, and $\sigma_r$ is the minimum transverse radius of the electron gas. Eq. (1) shows that at times $t = 2n\pi/\omega$ ($n = 0,1,2,...$) the transverse radius reaches its smallest value $\sigma_r$, and the electron number density concurrently maximizes. These periodic electron density peaks are responsible for the transient lensing effects in the probe images (*e.g.* Fig. 2).

Using Maxwell-Gauss law, we derive an expression for the radial electric field $\mathcal{E}_r(t)$ associated with the cylindrical Gaussian charge density (see SI2.2) and find $\mathcal{E}_r \propto r/\sigma_T^2$ in the limit $r < \sigma_T$, *i.e.* close to the center of the electron cloud the field is linear with radius $r$ which imparts the lensing effect on the probe electrons. Assuming that the duration of the interaction between the relativistic probe electrons and the electron gas is short compared to the evolution time scale of the gas, and also using the thin-lens approximation, we derive the focal length of the electron gas as (see SI2.3)

$$f_{EG}(t) = -\frac{2(2\pi)^{\frac{3}{2}}\epsilon_0 \gamma v_z^2 m_e}{Ne^2}\left[2\left(\frac{\sigma_v}{\omega}\right)^2 (1 - \cos(\omega t)) + \sigma_r^2\right] \quad , \tag{2}$$

where $\epsilon_0$ is the vacuum permittivity, $\gamma = 1.6$ is the relativistic Lorentz factor, $v_z = 2.3 \cdot 10^8$ m/s is the velocity of the 300 keV probing electrons, and $N$ is the number of electrons in the cloud. The focal length is inversely proportional to the number of electrons in the cloud, and it is negative as is expected for a diverging lens. Furthermore, it follows a similar periodic dependence as the transverse radius $\sigma_T(t)$, with minima in absolute focal length $|f_{EG}(t)|$ (maxima in lensing strength) occurring at times $t = 2n\pi/\omega$ ($n = 0,1,2,...$).

Eq. 2 shows that the number of electrons, the velocity spread, and the minimum transverse radius are needed to obtain values for the focal length of the electron gas. In the absence of any extracting field, the velocity spread of the electron cloud primarily comes from the photoemission process, Coulomb interactions at early times, and the dipole field between the electrons and their positive image charge[42,48,60-64]. The number of electrons in the electron gas is determined by the laser fluence and wavelength and by the absorption of electrons by the copper grid. Since these quantities are not known *a priori*, we estimate them from our data. In SI2.4 we demonstrate that the change in ROI intensity is inversely proportional to the electron-gas focal length when the intermediate lens is highly excited, *i.e.* for ILC > 1 A. This allows us to fit the ROI difference intensity trace for an ILC of 1.1 A to the function $1/f_{EG}(t) = Ae^{-t/\tau}/\sigma_T^2(t)$, where fitting parameter $A$ encompasses the number of cloud electrons and other unknown parameters that describe the strength of the lenses in the projection system of the TEM (see SI3 for fitting details). The exponential factor $e^{-t/\tau}$, with fitting parameter $\tau$, phenomenologically describes the slow decay of the electron lens strength due to the loss of electrons over time and dephasing. We set the minimum transverse radius to $\sigma_r = 12/\sqrt{2}$ μm, which is obtained from the experimental laser spot size of ~29 μm FWHM or $\sigma = 12$ μm (average of major and minor axes of elliptical footprint), and considering that the electrons are emitted through a two-photon process that scales quadratically with photon intensity. For concreteness, here we use a simplified model of the TEM post-sample lensing system allowing us to produce order-of-magnitude estimates of both $f_{EG}$ and the number of electrons $N \sim 10^5$ in the cloud (see SI2.4-5).

The resulting fit is shown in Fig. 7, together with the corresponding electron-gas focal length. Interestingly, the focal length magnitude varies by about a factor of ten, ranging from ~0.5 m at the lensing maxima to ~4-5 m in between cyclotron resonances. From the fit we are able to determine the angular frequency $\omega = 37.97 \pm 0.01$ GHz, and the transverse velocity spread $\sigma_v = 4.91 \pm 0.01 \cdot 10^5$ m/s (the standard deviations on the fit parameters do not reflect the inaccuracies of the model itself), the latter of which largely determines the width of the resonance peaks. The decay constant τ is on the order of several ns, but could not be determined with high accuracy due to the limited fitting window. We note that the quantities $\sigma_v$ and $\sigma_r$ are not expected to vary with time because the magnetic field does not do work and we are treating the self-interaction as negligible resulting in no appreciable electric field. Furthermore, the excellent agreement between the model and the data indicates that space-charge effects play a negligible role in this time regime (>100 ps). On the other hand, the amplitude of the ROI intensity change at the first peak predicted by theory is much larger than the experimentally measured amplitude. In fact, the ratio of the first to second peak amplitude is consistently ~0.3-0.5, regardless of the objective or intermediate lens strength or slight variations in the ROI radius (except for OLC = 0 A). Clearly, we need to include other elements such as Coulomb self-fields and the copper grid in order to quantitatively describe the early dynamics <100 ps.

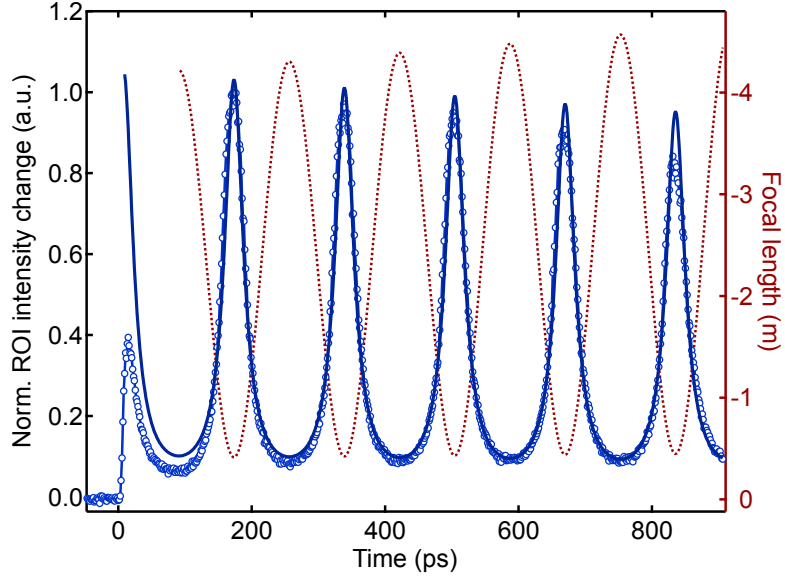

**Figure 7: Comparison between experimental and theoretical electron cyclotron dynamics.** The ROI difference intensity trace for ILC = 1.1 A (blue circles, left axis) is fitted to the function $1/f_{EG} = Ae^{-t/\tau}/\sigma_T^2(t)$ (dark blue line), where $A$ and $\tau$ are phenomenological fitting parameters encompassing the time-dependent number of electrons in the cloud and TEM-specific lens settings. $\sigma_T$ is the time-dependent transverse radius of the electron gas, which depends on the velocity spread $\sigma_v$, the minimum radius $\sigma_r$, and the cyclotron angular frequency $\omega$ (all fitted to the data). The fitting range is 100-900 ps, *i.e.* beyond the regime where Coulomb interactions are significant. The corresponding focal length of the electron gas, $f_{EG}$, is plotted in red (right axis), which is only an order-of-magnitude estimate due to the approximations in the model (see text).

**Numerical *N*-body simulations of the photoemitted electron cloud dynamics in the presence of Coulomb interactions and a copper grid.** In order to get a holistic picture of the electron dynamics, including Coulomb interactions within the electron cloud, we performed numerical *N*-body simulations for a realistic charge density that is subjected to a uniform, static magnetic field. Details of the numerical simulations are given in the Methods section. Briefly, with small time increments, we calculate the Lorentz force, $\boldsymbol{F} = q\boldsymbol{E} + q\boldsymbol{v} \times \boldsymbol{B}$, and resulting displacement of each electron inside an initially 3D Gaussian oblate ($2 \times 2 \times 0.01$ μm³) electron cloud with $10^4$ electrons. These parameters were chosen in order to approximately match the initial charge density in the experiment. The electromagnetic forces arise from self-Coulomb fields, the external magnetic field, as well as positive image charges due to the existence of the copper grid that acts as a planar conductor held at zero potential. We assume that electrons that hit the grid will be absorbed and hence omitted from the rest of the calculation. Snapshots taken from a simulation with $B = 0.22$ T at three time delays are superimposed in Fig. 8a (full movie S5). The frame at 16 ps after photoexcitation shows a flat electron distribution close to the copper grid, that has already significantly expanded due to Coulomb explosion of the gas during the first few ps. The distribution in the intermediate 83 ps frame, which corresponds to the first minimum in the electron cloud density, is homogeneously spread out over tens of μm in all $(x, y, z)$ directions. Finally, for the frame at 165 ps, which corresponds to the first cyclotron resonance peak, the electron gas regains its narrow transverse size, but it is severely elongated along the $z$-axis.

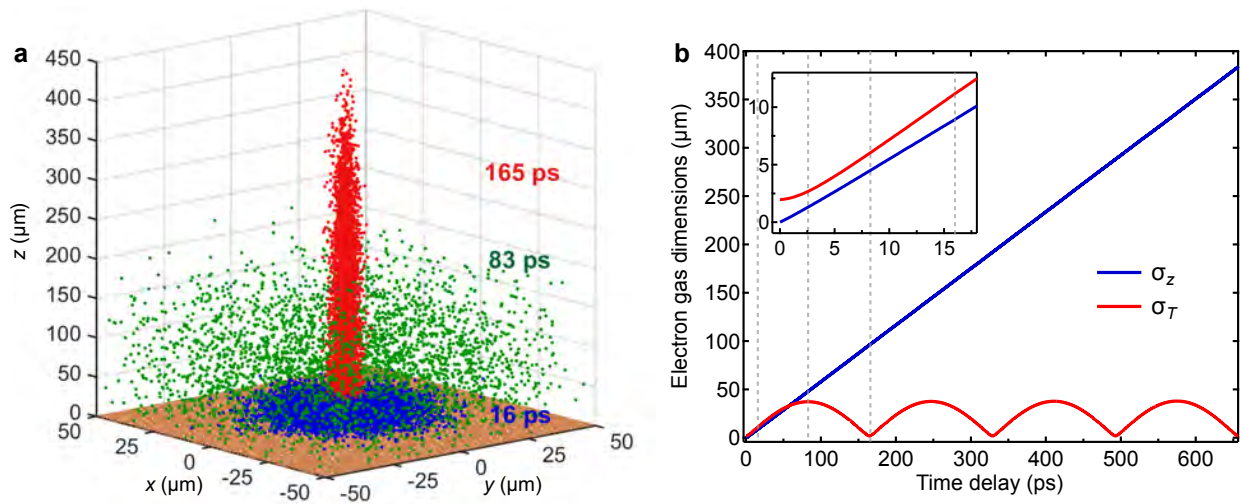

**Figure 8: Results from numerical *N*-body simulations. a** Snapshots of the electron distribution (10,000 electrons) taken at three different time delays. The copper grid is schematically shown in brown. **b** Electron gas dimensions (standard deviations) in the transverse ($x,y$) and axial ($z$) directions as a function of time, extracted from the *N*-body simulation with $B = 0.22$ T. The inset shows a zoom into the dynamics during the first 18 ps. The vertical lines indicate the time delays at which the snapshots in **a** were taken.

The corresponding radial and axial standard deviations of the electron cloud as a function of time are plotted in Fig. 8b. Within 165 ps, the electron gas morphs from an oblate (pancake-like) distribution, with an aspect ratio of $\sigma_z/\sigma_T \simeq 0.005$ and charge density of 40 mC/cm$^3$ (per $\sigma_T{}^2\sigma_z$), into a prolate (cigar-like) shape with an aspect ratio of $\sigma_z/\sigma_T \simeq 50$ and density of 2 μC/cm$^3$. This corresponds to a factor of $20 \cdot 10^3$ decrease in charge density, which is due to both the shape transformation as well as absorption of electrons by the grid (see SI4). For time delays >4 ps, the gas linearly expands along the $z$-direction with a velocity of $\sim 6 \cdot 10^5$ m/s, while it is refocused to a little past its initial radial size in the $x,y$-plane at intervals of $T = 165$ ps. At a time delay of 3 ns, the axial size of the electron gas reaches $\sim 1.5$ mm. The decay of the lensing signal and damping of the oscillations is therefore ascribed to a combination of continuing absorption of electrons by the copper grid (which is only $\sim 10\%$ over $\sim 6$ ns, see SI4), as well as a change of the cyclotron frequency due to the decrease of the axial magnetic field for distances >1 mm from the sample (see SI1). Absorption of electrons by the upper pole piece of the objective lens can also not be neglected on these time scales. Both latter effects are not taken into account in the simulation. The slow time scale of these loss/damping processes explains the exceptionally long lifetime of the cyclotron oscillations. Due to Coulomb explosion, the electron velocity distribution (mean, spread) changes abruptly during the first few ps after photoexcitation, but it reaches a plateau for time delays >50 ps (see SI5). The time-averaged transverse velocity spread from the simulation for times >50 ps, $\sim 7 \cdot 10^5$ m/s, is in reasonable agreement with the velocity spread obtained from fitting the analytical model to the ROI intensity data ($\sigma_v \sim 5 \cdot 10^5$ m/s, Fig. 7). This agreement confirms that the initial conditions of the simulations (electron density, velocity, and velocity spread) where chosen appropriately. The transverse velocity spread obtained from the simulation would match that of an equilibrated electron gas at a temperature of $\sim 3 \cdot 10^5$ K. Quantum phenomena such as Landau energy level quantization are therefore not expected to be observed experimentally.

The rapid oblate-to-prolate shape transformation of the electron gas, evidenced by the numerical simulations, has profound influence on its transient lensing strength. Indeed, the first peak in the ROI difference intensity traces is consistently lower in amplitude than the second peak. Qualitatively, we can attribute this to two things: first, the oblate shape of the electron gas at early times leads to a small transverse electric field component, and therefore a reduced impulse on the probing electrons. Second, the transverse electric field component is further reduced by the positive image charges, effectively creating a parallel-plate capacitor at early times. When the electron gas adopts a prolate shape elongated along the $z$-axis, the effect of the image charges is largely reduced since the electrostatic dipole force along $z$ scales with $\sim 1/d^2$ where $d$ is the distance between the two charges.

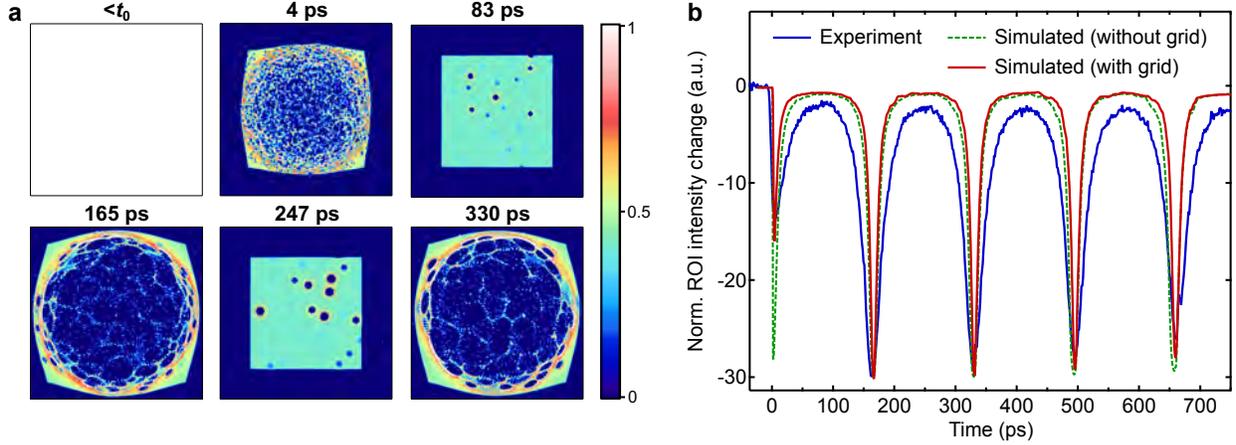

**Figure 9: Probe image simulations. a** Simulated grid images based on snapshots from the numerical *N*-body simulation. The sides of the images are 24 μm. Each electron on the detector is represented by a Gaussian kernel with a width of 3 pixels. **b** Simulated and experimental ROI difference intensity traces. The traces have been normalized to the second peak at 165 ps. The dashed green curve represents a calculation without a copper grid (no absorption of electrons, no image charges). The red curve includes the grid.

In order to confirm this interpretation on a more quantitative basis, we simulated the UEM lensing movies by sending a regular grid of relativistic probe electrons through each frame of the *N*-body simulation (see Methods section for details). Here, we neglect Coulomb interactions between probe electrons, as well as any perturbations of the electron gas by the probe electrons. Representative snapshots of these probe simulations are shown in Fig. 9a (full movie S6), which can be compared to the experimental movie frames in Fig. 2b. All features are reproduced well, including the depletion of the probe intensity in the center, a bright ring around the depletion area in the first peak at $\Delta t = 4$ ps (12 ps in the experiment), as well as a profound magnification of the grid images at the cyclotron resonance peaks ($\Delta t = 165$ ps, 330 ps). The lensing is much stronger at the cyclotron resonance peaks, when the electron gas adopts a prolate shape, than at the first peak, when it has an oblate shape. Corresponding ROI difference intensity traces, with and without the copper grid included, are shown in Fig. 9b, together with an experimental ROI trace taken at low IL currents. The agreement is satisfactory, in particular the ratio of the first and second peak amplitudes is reproduced very well, as well as the shape of the resonance peaks. The discrepancy in width of the peaks is assigned to differences in the electron velocity spread and the number of electrons, which are difficult to get right without explicitly including the photoemission process itself. We emphasize that the first-to-second peak amplitude ratio is only simulated well if the copper grid is included in the simulation. This shows that the image charges, as well as the absorption of electrons by the grid during the Coulomb explosion, play a significant role in the dynamics <50 ps.

CONCLUSIONS

Using a newly developed ultrafast electron microscope, we observed the ps-resolved cyclotron dynamics and lensing of a 3D hot electron gas created by photoemission from a copper target with intense fs laser pulses. Within 100-200 ps after photoexcitation, the gas undergoes an oblate-to-prolate shape transformation with a change in aspect ratio of a factor of $10^4$, and subsequent transverse expansions and contractions due to the gyration of individual electrons around the static magnetic field axis in the microscope. The cigar-shaped electron cloud acts as a diverging lens to the probe electrons, with focal lengths ranging from ~0.5-5 m during one cyclotron oscillation. We show that the observed lensing is dominated by a cooperative mean-field effect, as opposed to particle-particle scattering of individual probe and cloud electrons. Specifically, the granular nature of the electron distribution can effectively be ignored and instead can be replaced by the mean-field it creates (at least at the velocities we are considering here). Our current analytical treatment allows us to estimate the velocity spread and number of electrons in the gas, but it excludes the influence of Coulomb interactions inside the cloud, positive image charges, as well as the absorption of electrons by the grid. We performed numerical *N*-body simulations to take these effects into account, which proves to be crucial to understand and simulate the early dynamics before 50 ps. An analytical treatment including Coulomb interactions and image charges, and a more quantitative description of the TEM lensing system, will be part of future work.

These experiments inspire a plethora of future studies in at least three distinct fields. First, they present a unique way to directly visualize and characterize photoemitted charged-particle beams, which is of importance in the fields of high-brightness electrons sources for ultrafast microscopy and fourth-generation X-ray facilities, and plasma physics. Future experiments will focus on systematically investigating the dependence on laser wavelength (tuning the regime from two-photon, to one-photon and three-photon emission), and laser fluence. Furthermore, using an electrical TEM holder, one could apply a bias to the sample which enables studies below the virtual cathode limit[43,64]. Second, our work paves the way for the study of charge carrier cyclotron dynamics inside photoexcited materials using UEM. Such experiments would need to be performed at low temperatures, and need materials with large carrier diffusion lengths, such as InSb, InAs, GaAs/AlGaAs, or transition metal oxides[16,25]. Intense photoexcitation can create electron-hole plasmas, in which the electron and holes gyrate with different frequencies, direction, and spatial extent. Furthermore, the implementation of quantum point contacts using a custom MEMS-based TEM holder, could enable the spatiotemporal visualization of coherent flow and magnetic focusing of charge carriers in 2D electron-gas materials[31,34-36]. In this context, it is important to note that in conventional electron diffraction experiments in solids, the discrete structure of the particles in the target is extracted by taking a Fourier transform of the scattering data[65]. The local electric field in these systems is dominated by the local charge density, because the system is close to charge neutral. In contrast, in systems that are not charge neutral, such as charged particle bunches in free space, the long-range Coulomb force leads to local electric fields that are strongly influenced by all of the particles in the bunch. In this case the probe electron deflections are dominated by cooperative mean-field space-charge effects, and the scattering due to local charge inhomogeneities typical of scattering from solids is a second order effect. In considering cooperative lensing effects in electron systems confined in the solid state, such as at heterointerfaces or at surfaces, several factors arise; including scattering from the atomic structure of the hosting solid, and screening of the Coulomb interactions due to charge polarization in the solid. Though this modifies the lensing effects, especially on long length scales, qualitatively similar cooperative lensing effects may still be expected in cases where high density interfacial electron gases can be generated.

METHODS

**Ultrafast electron microscopy setup, experimental conditions, and data treatment.** We employ a custom-modified environmental *Hitachi H9500* TEM operating at 300 keV, interfaced with a high-repetition rate fs laser system (*Light Conversion PHAROS* with *ORPHEUS-F* OPA) that allows excitation of the sample with wavelengths between 260 and 2600 nm and variable repetition rates up to 1 MHz. In the experiments reported here, the sample is excited using 528 nm, ~200 fs laser pulses, with fluences of ~30 mJ/cm$^2$. Short probe electron pulses are generated using the photoelectric effect by impinging 256 nm, ~200 fs UV laser pulses onto a graphite guard-ring LaB$_6$ photocathode with a diameter of 50 μm (*Kimball Physics*). Laser pump and electron probe pulses impinge the sample with a repetition rate of 490 kHz and their relative delay is controlled using an optical delay line (*Aerotech*). The data acquisition software is provided by *IDES Inc.*

Typical integration times per image were 1 s, corresponding to $4.9 \cdot 10^5$ pump/probe shots. We note that the exact temporal resolution of the setup is not known yet. However, we excite the photocathode with a low pulse energy of 16 nJ, which puts us into a regime where tens-hundreds of electrons are emitted at the photocathode, and only a few electrons will reach the sample. This so-called "single-electron" mode has previously been shown to yield instrumental response functions (IRF) that are almost entirely limited by the pump and probe laser pulse durations[66], *i.e.* ~500 fs in our case. The rise of the ROI signal intensity itself cannot be used to determine the IRF, since our simulations indicate that the rise time is prolonged in the presence of image charges (see SI6). The sample consists of a 3000 mesh copper gilder grid *(SPI Supplies)* with a hexagonal hole diameter of ~4.5 μm, and thickness of ~5 μm. The overall transmission is 34%. The copper grid is tilted by 15° towards the pump laser beam in order to minimize the footprint to $22 \times 36$ μm FWHM. The angle between the pump laser and sample plane is 37°. Under these conditions we only expect photoemission from the top surface of the copper grid. Our laser fluences are below the ablation limit of copper[67]. Scanning electron microscopy images show no damage after prolonged laser exposure at ~20 mJ/cm$^2$ (see SI7).

Images are normalized to their total integrated intensity in order to compensate for slight variations in the probe electron intensity. A median filter of $5 \times 5$ pixels is applied to the images to mitigate random noise (the detector has $2000 \times 2000$ pixels). Difference images were generated by subtracting an averaged pre-time zero image from all subsequent frames. We also subtracted a frame recorded without probe electrons, but with pump laser beam in order to remove pump laser scatter that reaches the detector. Circular ROI radii were chosen with the goal of simultaneously optimizing the visibility and signal-to-noise ratio.

**Analytical model.** Our analytical derivations are based on the use of a Gaussian model for the charge distribution. For a non-interacting system it is a straightforward proof that the evolutions of the statistics of an ensemble of particles are independent of the spatial distribution of the particles; therefore, to most closely resemble the experimental conditions, we treat the spatial distribution as Gaussian in our analysis. The time-dependent density of the electron gas is described by

$$\rho(r, z, t) = \frac{-Ne}{(2\pi)^{\frac{3}{2}}} \frac{e^{-\frac{z^2}{2\sigma_z^2(t)}}}{\sigma_z(t)} \frac{e^{-\frac{r^2}{2\sigma_T^2(t)}}}{\sigma_T^2(t)} \ , \tag{3}$$

where $\sigma_T(t)$ is defined in Eq. (1) (see SI2 for derivation). The $N$-body simulation results indicate that $\sigma_z(t)$ becomes significantly larger than $\sigma_T(t)$ within 150 ps, which means at each time the charge distribution can be approximated by an infinitely long charged cylinder,

$$\rho(r, t) \approx \frac{-Ne}{(2\pi)^{\frac{3}{2}}} \frac{1}{\sigma_z(t)} \frac{e^{-\frac{r^2}{2\sigma_T^2(t)}}}{\sigma_T^2(t)} \ , \tag{4}$$

whose electric field can be obtained from Maxwell-Gauss law as (see SI2 for derivation)

$$\mathcal{E}_r(r, t) = \frac{-Ne}{(2\pi)^{\frac{3}{2}}\epsilon_0} \frac{1}{\sigma_z(t)} \frac{1 - e^{-\frac{r^2}{2\sigma_T^2(t)}}}{r} \ . \tag{5}$$

For $r < \sigma_T$

$$\mathcal{E}_r(r < \sigma_T, t) \approx \frac{-Ne}{2(2\pi)^{\frac{3}{2}}\epsilon_0} \frac{1}{\sigma_z(t)} \frac{r}{\sigma_T^2(t)} \ , \tag{6}$$

*i.e.* the transverse electric field is linear in the radial coordinate $r$. A linear electric field imparts a lensing effect on the probe electrons. Further derivations that lead to Eq. (2) are provided in the SI Section 2.

**Numerical simulations.** We start our *N*-body simulations with a very oblate ($2 \times 2 \times 0.01$ μm$^3$) 3D Gaussian slab containing $10^4$ electrons. The oblate electron slab is placed at a distance of 30 nm away from the copper surface before starting the simulations. The photoemission process itself is not included in this approach, but the rapid photoemission of the electrons renders the longitudinal dimension of the bunch very small, resulting in a pancake-like bunch after the photo-emission process is complete. We assume such a bunch for the initial conditions of our simulations. The initial velocity distribution is a Gaussian with a mean of $(0, 0, 6 \cdot 10^5$ m/s) in the $(x, y, z)$ directions and an isotropic spread with a standard deviation of $6 \cdot 10^5$ m/s in each axis. These values were chosen to approximately match the velocity spread obtained from the experimental data, as well as to avoid that all electrons are absorbed by the grid. Since the copper grid is grounded, the potential at its surface is zero. Therefore, there is an electric dipole field formed between the photoemitted electron gas and its positive image charge, which is mostly aligned parallel to the propagation direction of the probe electrons. Image charges are approximated by calculating the dipole field for each electron and its positive counter-charge located at the opposite site of the copper grid surface, *i.e.* the electron coordinates are mirrored at the sample plane to find the coordinates of the image charges. Finally, the copper grid is approximated as a plane surface neglecting the effects of the holes. The measured cyclotron frequencies (GHz) correspond to centimeter wavelengths, which is much larger than the hole size of the grid (~4 μm), which justifies the use of a homogeneous slab instead of a grid in the simulations. An electron that hits the conductive copper plate is absorbed and excluded from the simulation. The tilt of the grid is neglected. The Lorentz force $\boldsymbol{F} = q\boldsymbol{E} + q\boldsymbol{v} \times \boldsymbol{B}$ is calculated for each electron in the gas, and the equation of motion is solved using the finite difference method. A time step of 80 fs was chosen such that reducing its value further would not change the results considerably. It is assumed that the electrons move much slower than the speed of light and hence it is not necessary to use retarded electric fields or account for losses due to electromagnetic radiation.

The effect of the photoemitted electron gas on the probe electrons is simulated by placing 11025 electrons equally spaced on a square grid with 14 µm sides, centered on the optical axis, and starting their motion at $3\sigma_z(t)$ above the copper grid. The kinetic energy of the probe electrons is 300 keV, corresponding to a speed of $v_z = 2.3 \cdot 10^8$ m/s (or $0.77c$). The detector was placed at 1 m below the electron gas, and a ROI circle radius of 7 µm was used to plot the difference intensity traces. Except for the electron gas lens, no other lenses inside the TEM are considered. The comparison between experimental and simulated data is therefore only qualitative. We neglect Coulomb interactions between probe electrons, as well as any perturbations of the electron gas by the probe electrons.

SUPPLEMENTARY INFORMATION

**Movie S1**: Difference-image movie belonging to Fig. 2a

**Movie S2**: Full-image movie belonging to Fig. 2b

**Movie S3**: Full-image movie belonging to Fig. 4b

**Movie S4**: Full-image movie belonging to Fig. 4c

**Movie S5**: *N*-body 3D numerical simulation movie

**Movie S6**: *N*-body numerical simulation probe movie

**Section SI1**: Plot of OLC vs magnetic field including manufacturer data; plot of magnetic field as a function of *z*.

**Section SI2**: Details about the analytical treatment: (1) derivation of the equation for $\sigma_T(t)$ (Eq.1 in main text); (2) use of Maxwell-Gauss law to obtain electric field from charge density; (3) derivation of the focal length equation using linear-field approximation and impulse on probe electrons (Eq. 2 in main text); (4) derivation of relationship between $\Delta S_{ROI}$ and electron-gas focal length; (5) Determination of number of electrons from magnification of the grid image.

**Section SI3**: Details of the fitting procedure of $\Delta S_{ROI}$ data for ILC = 1.1 A and inverse of focal length, including all fitted parameters and errors.

**Section SI4**: Plot that shows absorption of electrons with and without grid and image charges

**Section SI5**: Plot that shows mean and spread of kinetic energies as a function of time

**Section SI6**: Plot that compares simulated ROI traces (no grid, grid no image charges, grid with image charges); zoom into early times to show rise time.

**Section SI7**: SEM images of copper grid after laser exposure.

ACKNOWLEDGEMENTS

This research was carried out in part at the Materials Research Laboratories Central Research Facilities, University of Illinois at Urbana-Champaign (UIUC). The authors thank Jim Mabon (UIUC) for assistance during the measurements, and Jian-Min Zuo (UIUC), Hyuk Park, and Ken Eberly (Hitachi High-Tech) for support during the early phases of instrument development. We thank Hitachi High-Tech for providing the data on the magnetic field strength of the objective lens. We thank Arthur Blackburn (University of Victoria) for useful discussions. Work by R.M.V., O.Z., A.E.S., R.D.C, and F.M.A. was in part supported by a Packard Fellowship in Science and Engineering from the David and Lucile Packard Foundation and by the National Science Foundation (NSF) CAREER award #1751725. R.M.V. gratefully acknowledges support from the Department of Chemistry at UIUC and John and Margaret Witt. Work by B.Z. and P.M.D. at Michigan State University is supported by the NSF though awards #1803719 and #1625181.

# Supplementary Information

# Transient lensing from a photoemitted electron gas imaged by ultrafast electron microscopy

*Omid Zandi[1,2], Allan E. Sykes[1,2], Ryan D. Cornelius[1,2], Francis M. Alcorn[1,2], Brandon Zerbe[4], Phillip M. Duxbury[4], Bryan W. Reed[5], Renske M. van der Veen[1,2,3]*

[1] Department of Chemistry, University of Illinois at Urbana-Champaign, Urbana, IL 61801, USA

[2] Materials Research Laboratory, University of Illinois at Urbana-Champaign, Urbana, IL 61801, USA

[3] Department of Materials Science and Engineering, University of Illinois at Urbana-Champaign, Urbana, IL 61801, USA

[4] Department of Physics and Astronomy, Michigan State University, East Lansing, MI 48824, USA

[5] Integrated Dynamic Electron Solutions, Inc. (IDES), Pleasanton, CA 94588, USA

# SI1 Magnetic field data from the manufacturer

In Fig. S1a we compare the magnetic field that is determined in this study using the cyclotron resonance frequency and the Larmor formula $B = 2\pi m_e/eT$, with the magnetic field for three different objective lens currents (OLC) obtained from the TEM manufacturer (courtesy *Hitachi High-Tech*). The data is fitted with linear functions that pass through (0,0). The slopes differ by ~5%, which is within the tolerance of magnetic field difference between the lens models and microscopes. Fig. S1b shows the magnetic field as a function of $z$-coordinate obtained from the manufacturer. The field is uniform for a region ±1 mm away from the eucentric height, where the sample is placed, at $z = 0$.

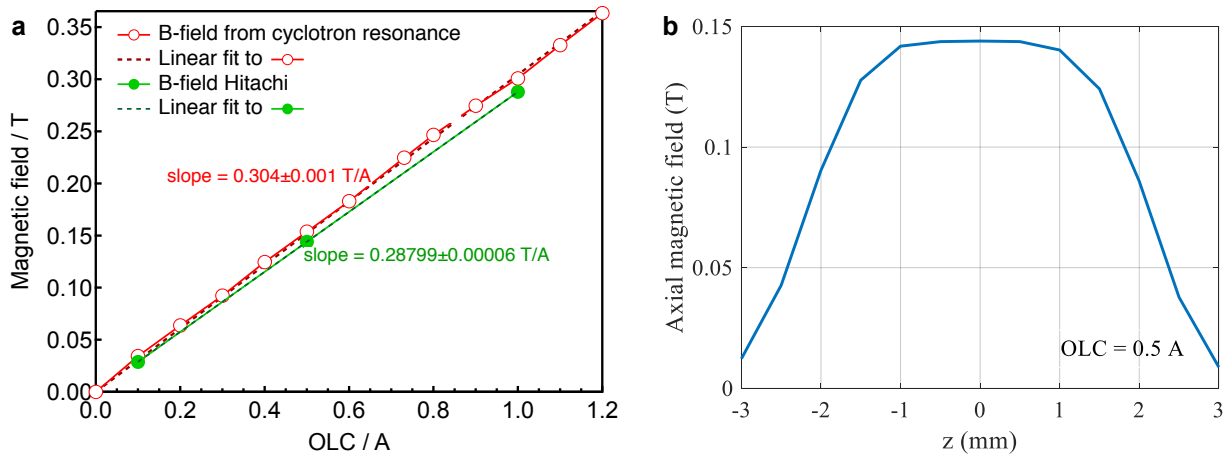

**Figure S1: a** Comparison between magnetic fields obtained from cyclotron resonances (red circles) and the ones from the manufacturer (courtesy *Hitachi High-Tech*) (green circles). The dashed lines are linear fits to the data. **b** Axial magnetic field as a function of z-coordinate (courtesy *Hitachi High-Tech*).

# SI2   Analytical derivations

**(1)** <u>**Derivation of the equation for $\sigma_T(t)$ (Eq. (1) in main text):**</u>

Consider a single electron created at time $t = 0$ at position $\mathbf{R} = x_0\hat{\mathbf{x}} + y_0\hat{\mathbf{y}} + z_0\hat{\mathbf{z}}$ with an initial velocity $\mathbf{v_0} = v_{x0}\hat{\mathbf{x}} + v_{y0}\hat{\mathbf{y}} + v_{z0}\hat{\mathbf{z}}$ inside a uniform time-invariant magnetic field of $\mathbf{B} = B_0\hat{\mathbf{z}}$ (Fig.

S2). Applying the Lorentz force, the momentum of the electron will be obtained from

$$\frac{d\mathbf{p}}{dt} = -e\,\mathbf{v} \times B_0\hat{\mathbf{z}} = -e(v_x\hat{\mathbf{x}} + v_y\hat{\mathbf{y}}) \times B_0\hat{\mathbf{z}} = eB_0(v_x\hat{\mathbf{y}} - v_y\hat{\mathbf{x}}). \tag{S1}$$

Assuming $\mathbf{p} = \gamma m\mathbf{v}$, we will have

$$v_x = v_0\cos(\omega t + \phi); \ \ v_y = v_0\sin(\omega t + \phi), \tag{S2}$$

where

$$\omega = \frac{eB_0}{\gamma m_e}; \ \ \ v_0\cos(\phi) = v_{x0}; \ v_0\sin(\phi) = v_{y0}. \tag{S3}$$

And the x and y coordinates of the electron will be

$$x = \frac{v_0}{\omega}\sin(\omega t + \phi) + x_0 - \frac{v_0}{\omega}\sin(\phi); \ \ y = -\frac{v_0}{\omega}\cos(\omega t + \phi) + \frac{v_0}{\omega}\cos(\phi) + y_0, \tag{S4}$$

the transverse distance between the electron at time $t$ from its initial position $\mathbf{R}$ will be

$$r(t) = \frac{v_0}{\omega}\sqrt{(\sin(\omega t + \phi) - \sin(\phi))^2 + (-\cos(\omega t + \phi) + \cos(\phi))^2} = \sqrt{2}\frac{v_0}{\omega}\sqrt{1 - \cos(\omega t)},$$

$$\tag{S5}$$

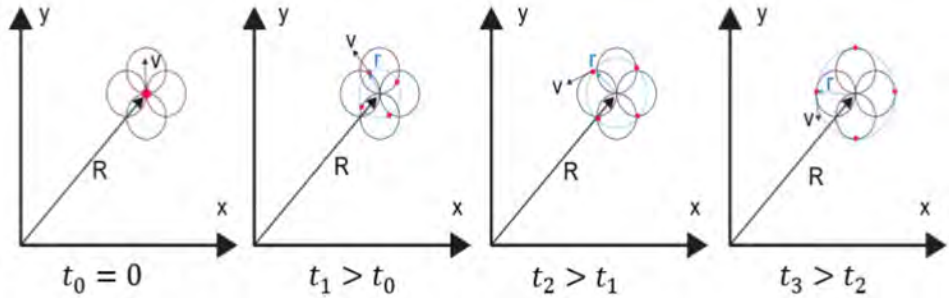

**Figure S2:** Position of four electrons at four different times. The initial velocity of the electrons have the same amplitude but has a uniform angular distribution.

Now we consider $N$ electrons as shown in Fig. S2 with a normal distribution in $z$. Then, if the initial electron velocity of the $N$ electrons has a uniform angular distribution, their charge density will be

$$\rho_{x_0,y_0,v_0}(x,y,z,t) = -Ne\frac{e^{-\frac{z^2}{2\sigma_z^2(t)}}}{(2\pi)^{1/2}\sigma_z(t)}\frac{\delta\left(\sqrt{(x-x_0)^2+(y-y_0)^2}-\sqrt{2}\frac{v_0}{\omega}\sqrt{1-\cos(\omega t)}\right)}{2\pi\sqrt{2}\frac{v_0}{\omega}\sqrt{1-\cos(\omega t)}},\tag{S6}$$

where $\delta(\cdot)$ is the Dirac delta function. Suppose the initial velocity amplitude has a zero-centered Gaussian distribution of

$$f_v(v_0) = \frac{1}{2\pi\sigma_v^2}e^{-\frac{v_0^2}{2\sigma_v^2}},\tag{S7}$$

with $\sigma_v$ being the standard deviation of the electrons velocity. Therefore, integration of Eq. (S6) over $f_v(v_0)$ gives

$$\rho_{x_0,y_0}(x,y,z,t) = \int\limits_0^\infty 2\pi v_0\,\rho_{x_0,y_0,v_0}(x,y,z,t)\,f_v(v_0)dv_0 =$$

$$= -Ne\frac{e^{-\frac{z^2}{2\sigma_z^2(t)}}}{(2\pi)^{3/2}\sigma_z(t)}\int\limits_0^\infty\frac{\delta\left(\dfrac{\sqrt{(x-x_0)^2+(y-y_0)^2}}{\frac{\sqrt{2}}{\omega}\sqrt{1-\cos(\omega t)}}-v_0\right)}{2\left(\frac{\sigma_v}{\omega}\right)^2(1-\cos(\omega t))}e^{-\frac{v_0^2}{2\sigma_v^2}}dv_0$$

$$= -Ne\frac{e^{-\frac{z^2}{2\sigma_z^2(t)}}}{(2\pi)^{3/2}\sigma_z(t)}\frac{e^{-\frac{(x-x_0)^2+(y-y_0)^2}{4\left(\frac{\sigma_v}{\omega}\right)^2(1-\cos(\omega t))}}}{2\left(\frac{\sigma_v}{\omega}\right)^2(1-\cos(\omega t))}.$$

$$\tag{S8}$$

The electrons have an initial spatial distribution of

$$f(x_0,y_0) = \frac{1}{(2\pi)^2\sigma_x\sigma_y}e^{-\frac{x_0^2}{2\sigma_x^2}-\frac{y_0^2}{2\sigma_y^2}},\tag{S9}$$

with $\sigma_x$ and $\sigma_y$ being the standard deviations of the zero-centered distribution in $x$ and $y$, respectively, which is determined by the laser pulse profile. Therefore, the charge density will be determined by

$$\rho(x,y,z,t) = \int\limits_{-\infty}^{\infty} \int\limits_{-\infty}^{\infty} \rho_{x_0,y_0}(x,y,z,t) f(x_0,y_0) dx_0 dy_0$$

$$= -Ne \frac{e^{-\frac{z^2}{2\sigma_z^2(t)}}}{(2\pi)^{3/2}\sigma_z(t)} \frac{1}{2\left(\frac{\sigma_v}{\omega}\right)^2(1-\cos(\omega t))} \frac{1}{(2\pi)^2 \sigma_x \sigma_y} \int\limits_{-\infty}^{\infty} \int\limits_{-\infty}^{\infty} e^{-\frac{(x-x_0)^2+(y-y_0)^2}{4\left(\frac{\sigma_v}{\omega}\right)^2(1-\cos(\omega t))}} e^{-\frac{x_0^2}{2\sigma_x^2} - \frac{y_0^2}{2\sigma_y^2}} dx_0 dy_0$$

$$= -Ne \frac{e^{-\frac{z^2}{2\sigma_z^2(t)}}}{(2\pi)^{1/2}\sigma_z(t)} \frac{1}{2\left(\frac{\sigma_v}{\omega}\right)^2(1-\cos(\omega t))} \frac{1}{(2\pi)^2 \sigma_x \sigma_y}$$

$$\times \int\limits_{-\infty}^{\infty} e^{-\frac{(x-x_0)^2}{4\left(\frac{\sigma_v}{\omega}\right)^2(1-\cos(\omega t))}} e^{-\frac{x_0^2}{2\sigma_x^2}} dx_0 \int\limits_{-\infty}^{\infty} e^{-\frac{(y-y_0)^2}{4\left(\frac{\sigma_v}{\omega}\right)^2(1-\cos(\omega t))}} e^{-\frac{y_0^2}{2\sigma_y^2}} dy_0 \, .$$

$$(S10)$$

From the identity

$$\int_{-\infty}^{\infty} e^{-(x-x0)^2/A} e^{-(x)^2/B} \, dx = \frac{e^{-\frac{x0^2}{A+B}}\sqrt{\pi}}{\sqrt{\frac{1}{A}+\frac{1}{B}}},$$

$$(S11)$$

Eq. (S10) becomes

$$\rho(x,y,z,t) = \frac{-Ne}{(2\pi)^{\frac{3}{2}}} \frac{1}{4\left(\frac{\sigma_v}{\omega}\right)^2(1-\cos(\omega t))} \frac{e^{-\frac{z^2}{2\sigma_z^2(t)}}}{\sigma_x \sigma_y \sigma_z(t)}$$

$$\times \frac{e^{-\frac{x^2}{4\left(\frac{\sigma_v}{\omega}\right)^2(1-\cos(\omega t))+2\sigma_x^2}}}{\sqrt{\frac{1}{4\left(\frac{\sigma_v}{\omega}\right)^2(1-\cos(\omega t))}+\frac{1}{2\sigma_x^2}}} \frac{e^{-\frac{y^2}{4\left(\frac{\sigma_v}{\omega}\right)^2(1-\cos(\omega t))+2\sigma_y^2}}}{\sqrt{\frac{1}{4\left(\frac{\sigma_v}{\omega}\right)^2(1-\cos(\omega t))}+\frac{1}{2\sigma_y^2}}} \, .$$

If $\sigma_v/\omega > \sigma_x, \sigma_y$, the transverse profile of the electron cloud will be mostly determined by the electron velocity distribution rather than the laser profile. Therefore, for the sake of simplicity, we assume that $\sigma_x = \sigma_y = \sigma_r$ and for $r^2 = x^2 + y^2$, we find

$$\rho(r,z,t) = \frac{-Ne}{(2\pi)^{\frac{3}{2}}} \frac{e^{-\frac{z^2}{2\sigma_z^2(t)}}}{\sigma_z(t)} \frac{e^{-\frac{r^2}{2\sigma_T^2(t)}}}{\sigma_T^2(t)} , \qquad (S13)$$

where

$$\sigma_T(t) = \sqrt{2\left(\frac{\sigma_v}{\omega}\right)^2 (1 - \cos(\omega t)) + \sigma_r^2} , \qquad (S14)$$

and $\sigma_z(t)$ increases linearly by time after the Coulomb explosion regime. Eq. (S14) is the same as Eq. (1) in the main text.

## (2) Use of the Maxwell-Gauss law to obtain the electric field from the charge density:

When $\sigma_z(t) \gg \sigma_T(t)$ the charge distribution can be approximated by an infinitely long charged cylinder, and we may write

$$\rho(r,t) \approx \frac{-Ne}{(2\pi)^{\frac{3}{2}}} \frac{1}{\sigma_z(t)} \frac{e^{-\frac{r^2}{2\sigma_T^2(t)}}}{\sigma_T^2(t)} , \qquad (S15)$$

whose electric field can be obtained from the Maxwell-Gauss law by

$$\mathcal{E}_r(r,t) = \frac{-Ne}{(2\pi)^{\frac{3}{2}}\epsilon_0} \frac{1}{\sigma_z(t)} \frac{1 - e^{-\frac{r^2}{2\sigma_T^2(t)}}}{r} . \qquad (S16)$$

For $r < \sigma_T$, the first term of the Taylor expansion of Eq. (S16) is

$$\mathcal{E}_r(r < \sigma_T, t) \approx \frac{-Ne}{2(2\pi)^{\frac{3}{2}}\epsilon_0} \frac{1}{\sigma_z(t)} \frac{r}{\sigma_T^2(t)} . \qquad (S17)$$

## (3) Derivation of the focal length equation using the linear-field approximation and impulse on probe electrons (Eq. (2) in main text):

The electron probe equation of motion in the transverse direction is

$$\frac{dp_r}{dt} = -e\mathcal{E}_r(r,t) \quad \text{for} \quad |t - t_0| < \frac{\tau_p}{2} \ , \tag{S18}$$

where $p_r$ is the transverse momentum of the probe electron, $t_0$ is the time the probe electron arrives at the sample plane, $\mathcal{E}_r$ is the transverse component of the electric field and $\tau_p$ is the time that the probe electron traverses the sample area and can be approximated by

$$\tau_p = \frac{\sigma_z(t)}{v_z} \ , \tag{S19}$$

where $\sigma_z(t)$ is the std of the electron cloud in the z direction and $v_z$ the probe electron velocity. Solving Eq. (S18) gives

$$\Delta p_r = p_r - p_{r0} = -e \int_{t_0 - \frac{\tau_0}{2}}^{t_0 + \frac{\tau_0}{2}} \mathcal{E}_r(r,t)dt \approx -e\frac{\sigma_z(t)}{v_z}\mathcal{E}_r(r,t_0) \ , \tag{S20}$$

where $p_{r0}$ is the initial momentum of the electrons and indeed represents the divergence of the probe electron beam determined by its brightness (at the TEM condenser stage). For an almost parallel beam, we set $p_{r0} = 0$. Solving for transverse displacement, we have

$$\Delta r = r - \frac{e\sigma_z(t_0)}{\gamma v_z m_e}\mathcal{E}_r(r,t_0)(t - t_0) = r - \frac{e\sigma_z(t)}{\gamma v_z^2 m_e}\mathcal{E}_r(r,t_0)(z - z_0) \ , \tag{S21}$$

where, as is shown in Fig. (S3), $\Delta r$ is the time-dependent transverse distance of the probe electrons from the optical axis, and $r$ is the initial transverse distance. Inserting the electric field from Eq. (S17) into Eq. (S21) gives

$$\Delta r = r - \frac{e\sigma_z(t_0)}{\gamma v_z m_e}\mathcal{E}_r(r,t_0)(t - t_0) = r + \frac{Ne^2}{2(2\pi)^{\frac{3}{2}}\epsilon_0\gamma v_z^2 m_e}\frac{r}{\sigma_T^2(t)}(z - z_0) \ . \tag{S22}$$

Solving Eq. (S22) for $\Delta r = 0$ gives the focal length of the electron gas as

$$f_{EG} = -\frac{2(2\pi)^{\frac{3}{2}}\gamma\epsilon_0 v_z^2 m_e}{Ne^2}\sigma_T^2(t) \tag{S23}$$

which is Eq. (2) in the main text.

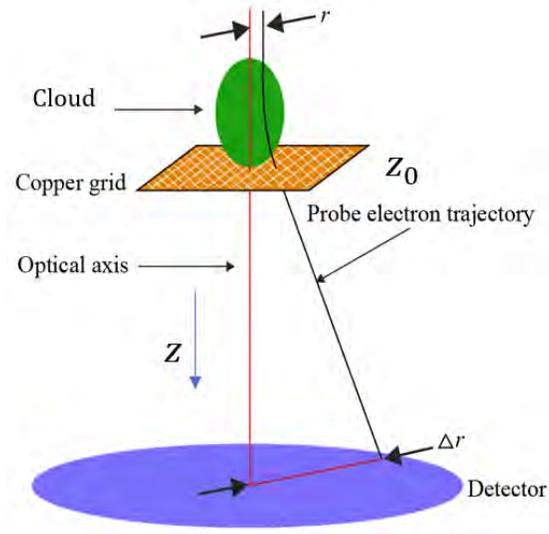

**Figure S3**: Deflection of the probe electrons by the electron gas.

## (4) Derivation of relationship between $\Delta S_{ROI}$ and electron-gas focal length:

Suppose we replace all the lenses after the cloud (sample plane) by a single lens whose focal length is $f_{EL}$ and we call it the equivalent lens (EL). In the thin lens approximation, the total focal length of the lensing system (equivalent lens and electron-gas lens) is

$$f_{total}(t) = \frac{f_{EL}f_{EG}(t)}{f_{EL} + f_{EG}(t) - d} \tag{S24}$$

where $d$ is the distance between the cloud and the EL. We assume $d$ (cm) is negligible in comparison to other dimensions (m). Fig. S4 shows the simplified imaging system where probe electrons with the initial radius $r_i$ pass through the lens and hit the detector.

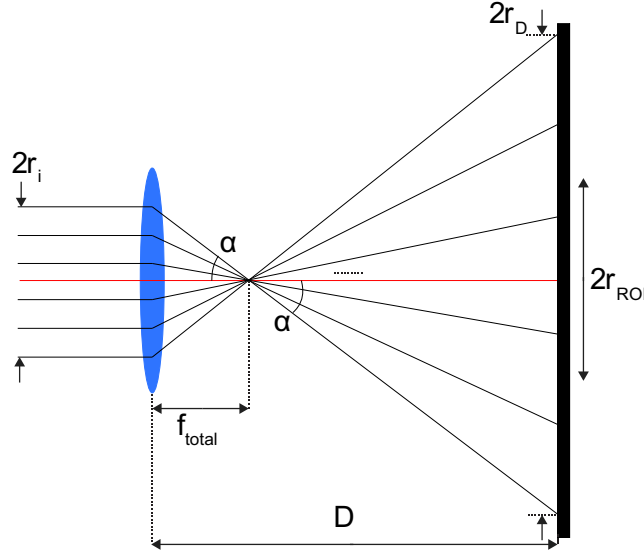

**Figure S4:** A simplified model of the lensing system that represents the cloud lens plus a lens equivalent to all the TEM lenses after the sample.

According to Fig. S4

$$\tan \alpha = \frac{r_i}{f_{total}} = \frac{r_D}{D - f_{total}} \rightarrow r_D = \frac{r_i}{f_{total}}(D - f_{total}) \tag{S25}$$

where $r_D$ is the radius of the probe beam on the detector and $D$ is the distance between the lensing system and the detector. The density of the probe electrons on the detector is

$$\rho_D = \frac{N_p}{\pi r_D^2} \ , \tag{S26}$$

where $N_p$ is the number of probe electrons. The detected intensity in the ROI is

$$S_{ROI} = 2\pi \int_0^{r_{ROI}} \rho_D r dr = N_p \frac{r_{ROI}^2}{r_D^2} = N_p \frac{r_{ROI}^2}{r_i^2 (D - f_{total})^2} f_{total}^2 \ . \tag{S27}$$

By assuming $D \gg f_{total}$, for a high excitation of the intermediate lens (>1 A), we have

$$S_{ROI} \approx \frac{N_p r_{ROI}^2}{r_i^2 D^2} f_{total}^2 \ . \tag{S28}$$

If we subtract the pre-time zero signal, we will get

$$\Delta S_{ROI} \approx N_p \frac{R_{ROI}^2}{r^2 D^2} (f_{total}^2 - f_{EL}^2) \ . \tag{S29}$$

Then

$$\Delta S_{ROI} \approx N_p \frac{R_{ROI}^2}{r^2 D^2} \left( \frac{f_{EL} f_{EG}(t)}{f_{EL} + f_{EG}(t) - d} + f_{EL} \right) \left( \frac{f_{EL} f_{EG}(t)}{f_{EL} + f_{EG}(t) - d} - f_{EL} \right)$$

$$\approx -N_p \frac{R_{ROI}^2 f_{EL}^2}{r^2 D^2} \left( 2 - \frac{f_{EL} - d}{f_{EG}(t)} \right) \left( 1 - \frac{f_{EL} - d}{f_{EG}(t)} \right) \frac{f_{EL} - d}{f_{EG}(t)} \approx -2N_p \frac{R_{ROI}^2 f_{EL}^2}{r^2 D^2} \frac{f_{EL} - d}{f_{EG}(t)},$$

$$(S30)$$

where we have assumed $f_{EG}(t) \gg f_{EL} - d$ (*i.e.* the electron-gas lens is much weaker than the TEM projection lenses). Therefore, if we excite the intermediate lens strongly, while the effect of the cloud lensing is still observable, the detected signal in the region of interest is approximately inversely proportional to the focal length of the cloud.

We note that this description of the lens system is highly simplified. It serves as a phenomenological model that gives us the right scaling in the focal length and the number of electrons (see below). Future work will focus on making this treatment more quantitative by taking into account the divergence of the incoming electron beam, and the distances between and excitation of all TEM lenses. In addition, geometric effects, such as the rather large dimension of the electron cloud along the $z$-direction, also need to be taken into account in order to reach quantitative agreement.

### (5) Estimation of the number of electrons in the cloud:

From Eq.'s (S16) and (S17), the cloud radial electric field maximizes when $\sigma_T(t)$ is minimum at $\omega t = 2n\pi; n = 0,1,2, ...$ for which the amount of probe electron deflection by the cloud depends only on $N$ and its minimum transverse size $\sigma_r$. From Eq. (S22), the number of electrons is

$$N = 2(2\pi)^{\frac{3}{2}} \epsilon_0 \gamma v_z^2 \frac{m}{e^2} \frac{\sigma_r^2}{(z_d - z_0)} \left( \frac{\Delta r}{r} \Big|_{\omega t = 2n\pi} - 1 \right) \qquad (S33)$$

where $\left.\frac{\Delta r}{r}\right|_{\omega t = 2n\pi}$ can be regarded as the highest magnification caused by the cloud and can be calculated as is shown in Fig. S5 and $z_d - z_0 = 100$ cm is the approximate distance between the cloud and the detector. The number of electrons then becomes $\sim 2 \times 10^5$. Because of all the approximations, we expect this method to give us the correct order of magnitude in the number of generated electrons.

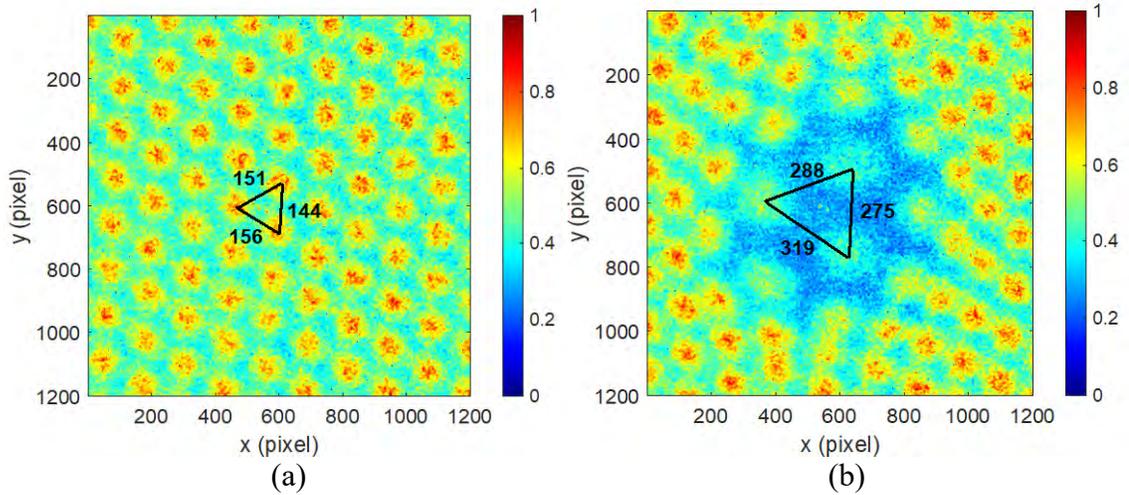

**Figure S5: a** Pre-time-zero image, and **b** an image at the second peak (first cyclotron resonance). A maximum magnification of $\sim 2$ is derived. The center of the grid holes are found by fitting Gaussians.

## SI3   Details of the fit of the ROI difference intensity trace

Eq. (S30) suggest that we can fit the inverse of the focal length function to the ROI signal in the focused regime with high excitation of the IL (current 1.1 A). Accordingly, we fit the function

$$1/f_{EG}(t) = \frac{A\,e^{-(t-t_0)/\tau}}{2\left(\frac{R}{\omega}\right)^2(1-\cos(\omega(t-t_0)))+1} \tag{S31}$$

to the ROI signal as is done in Fig. 7 in the main text. Here, $A$ is a constant that comes from combining Eq. (S23) with Eq. (30), $\tau$ is the damping time due to the absorption of the electrons and dephasing, $t_0$ is an arbitrary initial time (or $\omega t_0$ is the phase of oscillation), $\omega$ is the cyclotron angular frequency, $\sigma_v$ is the velocity spread and $\sigma_r$ is the minimum transverse size of the cloud (all σ denote standard deviations). The amplitude $A$ and $(\sigma_v, \sigma_r)$ are directly correlated with each other and therefore they cannot be determined independently. We therefore fit the ratio $R = \sigma_v/\sigma_r$, and set $\sigma_r = 12/\sqrt{2}$ μm, which is obtained from the experimental laser spot size of ~29 μm FWHM or σ = 12 μm (average of major and minor axes of elliptical footprint), and considering that the electrons are emitted through a two-photon process that scales quadratically with photon intensity. Using a non-linear least-square fitting procedure (built in the Matlab curve-fitting toolbox) for time delays 100-900 ps (*i.e.* passed the Coulomb explosion regime) we then obtain: $A = 1.05 \pm 0.01$ m$^{-1}$, $\omega = 37.97 \pm 0.01$ GHz, $\sigma_v = 4.91 \pm 0.01 \cdot 10^5$ m/s, $t_0 = 7.8 \pm 0.1$ ps, and $\tau = 8.2 \pm 0.2$ ns.

## SI4   Simulated absorption of electrons by the copper grid

In Fig. S6 we plot the number of electrons in the electron cloud as a function of time. The simulation starts with $10^4$ electrons whose center is placed at a distance of 30 nm away from the copper grid. During the first few ps, the electron gas undergoes a Coulomb explosion due to the large density and electron-electron repulsion. This leads to a large fraction of electrons being absorbed by the grid. As expected, more electrons are absorbed when image charges are included in the simulation. At later times, the electron absorption rate decreases until the fraction of electrons left in the simulation levels off at ~50%.

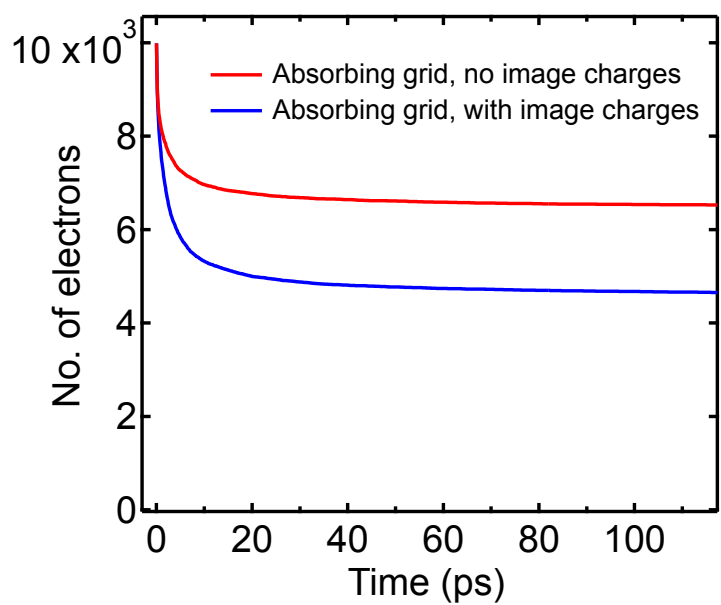

**Figure S6:** Number of electrons in the electron gas as a function of time delay extracted from a simulation with and without image charges. A significant fraction of electrons is absorbed by the copper grid.

## SI5   Mean kinetic energy and kinetic energy spread from simulations

Fig. S7 shows the mean kinetic energy of all electrons in the gas (**a**) and the kinetic energy spread (standard deviation) in **b** (with absorbing grid and image charges). For very early times (<1 ps), the mean kinetic energy and its spread spike to very large values (not shown) due to acceleration of the electrons towards the grid; the *y*-axis scale has been cut in order to show the data at later times. A plateau is reached around 50 ps after excitation, which indicates the regime where Coulomb interactions do not play a large role anymore.

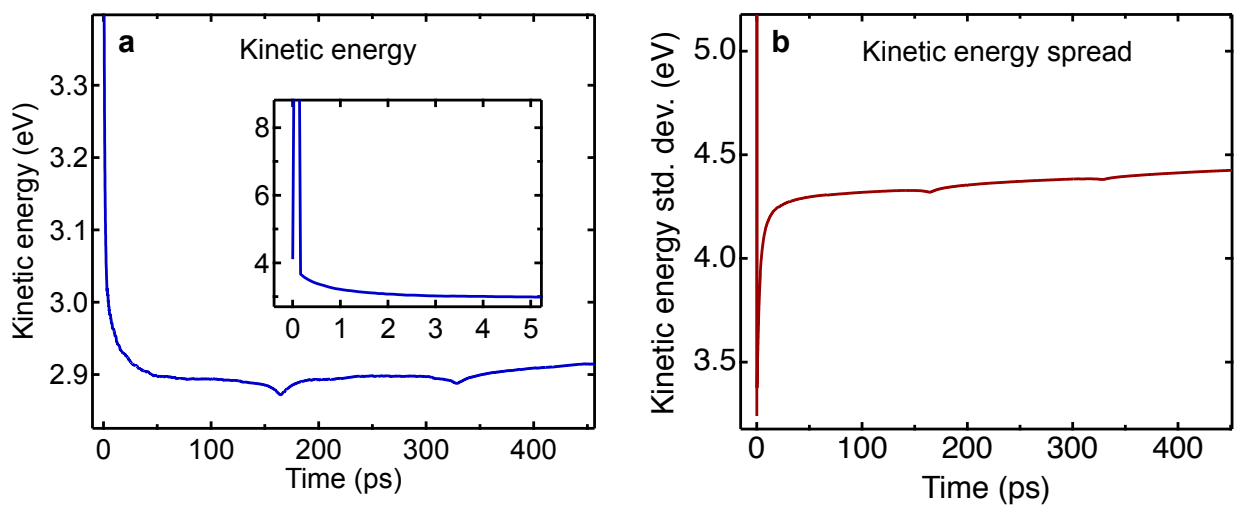

**Figure S7: a** Mean kinetic energy extracted from the simulation data. The inset shows a zoom into the early dynamics (0-5 ps). **b** Kinetic energy spread (standard deviation) energy extracted from the simulation data. Note that a fraction of electrons is absorbed by the copper grid (see Fig. S6).

# SI6 Simulations of ROI intensity traces

In Fig. S8 we show a set of simulated ROI difference intensity traces extracted from *N*-body probe simulations (see main text for details) for three cases: (1) without a grid (no absorption of electrons, no image charges); (2) with an absorbing grid, but without image charges; (3) with grid and image charges. It is clear that the majority of the amplitude reduction of the first peak is coming from the image-charge effect, which includes the increased absorption of electrons due to the dipole field between electrons and image charges (see Fig. S6). Fig. S8b shows a zoom into the first tens of ps after photoexcitation. It is seen that the rise time of the ROI intensity depletion signal is prolonged in the case of image charges, even though the creation process of the electron cloud was not explicitly included in the simulation. The latter could prolong the rise time even further.

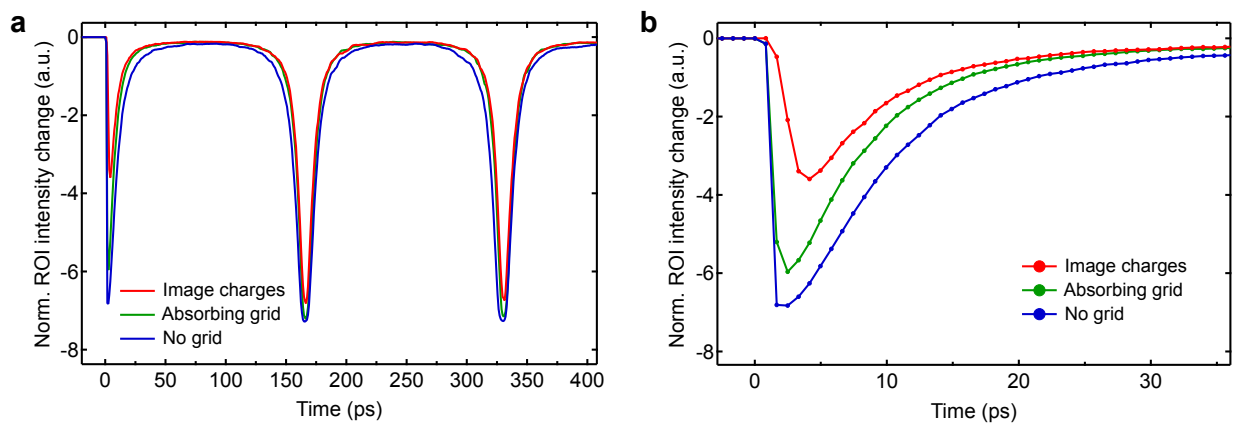

**Figure S8: a** Simulations of the ROI difference intensity obtained from the *N*-body simulations. **b** Same data as in **a**, but zoomed into the first 35 ps.

# SI7    SEM images of copper grid

Fig. S9 shows scanning electron microscopy (SEM) images of a 200 mesh copper grid, one taken before laser exposure and one taken after ~8 h of exposure at 528 nm, ~200 fs, ~20 mJ/cm². No damage is seen within the resolution of the images.

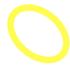

**Figure S9:** SEM images if a 200 mesh copper grid before (left) and after (right) laser exposure (528 nm, ~200 fs, 20 mJ/cm²). The approximate irradiation area is indicated by a yellow ellipse in the left figure.